# Controlling the LSPR properties of Au triangular nanoprisms and nanoboxes by geometrical parameter: a numerical investigation

Jagmeet Singh Sekhon[a,b]* and S S Verma[b]
[b]Department of Physical Sciences,
Indian Institute of Science Education & Research (IISER) Mohali, (Punjab), India 140306
[b]Department of Physics,
Sant Longowal Institute of Engineering & Technology (SLIET) Longowal,
District- Sangrur (Punjab), India 148106
*Corresponding Author emailed: jagmeetsekhon@ymail.com

**Abstract**

We have simulated the extinction spectra of Au triangular nanoprisms and nanoboxes by finite difference time domain (FDTD) method. It is found that the refractive index sensitivity increases linearly and near exponentially as the aspect ratio of nanoprisms increases and wall thickness of nanoboxes decreases. A sensing figure of merit (FOM) calculations shows that there is an optimum wall thickness for each edge length and height of the box, which makes them to be promising candidate for effective sensing applications. We have also shown that the higher FOM in triangular nanoboxes compared to the cubic nanoboxes and other solid structure is inherent in the shape of nanoparticles.

**INTRODUCTION**

The true utility of the metal nanoparticles is the ability to alter the localized surface plasmon resonance (LSPR) properties by engineering their structures. The structure change from spherical to sharp edge and solid to hollow, results the red-shift in LSPR wavelength [1-8]. Experimental and numerical analysis of nanometer sized hollow non-spherical structures [9-12] of diverse shapes focus the current area of plasmonics for their applications towards the optical sensing. Studies on these sharp edge solid or hollow nanostructures for better sensitivity to the environmental change are still to be explored for their optimum use. A number of Au nanoparticles have been employed in a variety of configurations to improve the detection limits with regards to the effect of nanoparticle geometry. [3-5] Schatz group from their studies on silver nanoprisms demonstrated that the LSPR properties can be significantly altered by height and edge length of the particle [7-8]. Cao et al. [3] investigated the optimum wall thickness for a fixed inner edge length of Au cubic nanoboxes. To the best of our knowledge, controlling the LSPR properties by geometrical parameters has never been reported for Au triangular nanoprisms and nanoboxes to find their optimum response towards the environmental refractive index change.

In this article, using FDTD software from Lumerical simulation [13], we have investigated the dependence of the LSPR wavelength and refractive index sensitivity on the edge length and height of triangular nanoprisms and also of inner edge length, wall thickness, and height of Au triangular nanoboxes, with the aim to find an optimum configuration for effective LSPR based biological sensing.

**RESULTS AND DISCUSSIONS**

The three dimensional theoretical analysis of the extinction spectra of triangular nanoprisms and nanoboxes of Au are done using the FDTD software with total field/scattered field (TFSF) plane wave source that is especially useful for single nanostructure. The direction of the light electromagnetic radiation is taken to be perpendicular to the plane of structure. All the calculations presented here refer to the embedded medium refractive indices of 1.00, 1.33, 1.35, and 1.37.

**Triangular nanoprisms**

The extinction cross-section of the gold triangular nanoprisms of various edge lengths (L) and heights (h) is depicted in Figure 1. The extinction intensity increases with both the edge length and height. The LSPR peak position shift towards the blue end as the nanoprism height increases for fixed edge length and also as the edge length decreases for fixed height whereas, the reverse change in shape parameters makes the shift towards red. The observed blue-shift is more dominating in case of large edge length. Another effect that comes into play is that the blue-shift with the decrease in edge length is more dominating in small height nanoprisms. Hence, the LSPR spectra can be easily tuned over the visible to near infrared regime by controlling the geometrical parameters of the nanoprism. Interestingly, the tunability of LSPR is not limited in the triangular nanoprisms as in case of nanospheres [14] and nanocubes [15] but it is like that of nanorods [14].

To understand the effect of edge length and height on the LSPR in a more simplified manner, we have plotted the LSPR wavelength against the aspect ratio (R, ratio of edge length to height) and illustrated in Figure 2A that shows the linear dependence of LSPR with aspect ratio. The linear fit to the calculated data point is $\lambda_{LSPR} = 608.36 + 34.24R$ in surrounding medium refractive index of 1.33. Hence, the sensing parameters are further plotted against the aspect ratio in case of triangular nanoprisms. To determine the response of the LSPR spectra towards the bulk refractive index change, we have calculated the refractive index sensitivity (RIS) defined as the relative changes in LSPR wavelength with respect to a change in the refractive index of the surrounding medium, $d\lambda_{LSPR}/dn$ and pictured in Figure 3A. Interestingly, the RIS enhances linearly with the aspect ratio, same as in case of Au nanorods [14]. The sensitivity of is higher than that of Au nanospheres [14] and nanocubes [15]. Again the enhancement in RIS is linear with the aspect ratio and expression from the linear fit is $RIS\,(nm/RIU) = 283.91 + 29.45R$.

Figure of merit (FOM, ratio of refractive index sensitivity to resonant line-width) is also calculated to optimize the aspect ratio and LSPR spectral region for better sensing response. The optimum aspect ratios are about 3.5 and 7.0 where FOM is about 12.5 and 12.6, respectively (Figure 2B). Now to optimize the LSPR spectral regime, both RIS and FOM are plotted against the resonance (LSPR) wavelength in Figure 3. From the FOM viewpoint, the optimum sensing region is 722 nm and 876 nm. Figure 3 is an evidence for the linear behaviour of RIS with the resonance wavelength and shows the shape independence of the sensitivity towards surrounding environment and the linear fit equation is $RIS(nm/RIU) = -299.46 + 0.85\lambda_{LSPR}$. This expression is an evidence to show that the sensitivity of nanoparticle plasmon resonance towards the environmental change can be directly obtained from the resonance peak position instead of going into detail of the nanoparticle

structure as also shown form the studies on nanorods [16] and nanocubes [15].

**Triangular nanoboxes**

Based on electrodynamic approach, Cao et al. [3] have demonstrated that the Au nanoboxes of cube shape show the higher RIS in comparison with the solid nanostructures i.e., nanocubes. Present work shows the interesting results for Au nanoboxes of triangular shape based on the FDTD simulations. It is well known that by change in the wall thickness of the nanoboxes, the plasmonic characteristics can be significantly altered [3]. The calculated extinction spectra of the Au triangular nanoboxes have been presented for inner edge length of 51.9 nm and inner height of 15 nm in Figure 4. The optical spectra shows the expected blue-shift in peak position as the wall thickness increases for fixed inner edge length as in case of cubic nanoboxes and shown as example in Figure 4A. Another interesting point is that as the wall thickness increases the peak intensity also increases and the second peak in extinction spectra starts disappearing. The nanoboxes optical extinction spectra shift towards the near infrared regime as the inner edge length increases and the inner height decreases (Figure 4B). The peak intensity increases as the height and inner edge length increases. Hence, the shape (i.e., height, resonant line-width, and resonance peak position) of extinction spectra shows direct dependence on the edge length, height, and wall thickness of Au nanoboxes. Figure 4 clearly illustrates that the LSPR wavelengths can be easily tuned by controlling the wall thickness, inner edge length, and height of the nanoboxes.

The refractive index sensitivity and FOM are again calculated to see the effect of change in structure from solid to hollow on their comparative sensing performance. The RIS rises near-exponentially with the decrease in wall thickness (Figure 5A). This trend is very similar to that of the cubic nanoboxes reported by Cao et al. [3]. The maximum of refractive index sensitivity is obtained for the edge length of 34.6 nm and 51.9 nm with the box height of 15 nm and wall thickness of 2.5 nm. Interestingly, the obtained RIS is much higher for the triangular nanoboxes in comparison with the cubic nanoboxes [3] and nanoprisms of similar dimensions. An interesting result comes out when FOM is calculated and plotted against the wall thickness $(t)$ in Figure 5B. It can be seen that, with decreasing $t$ from 10 nm to 2.5 nm, FOM increases slowly first, then reaches maximum value, and finally decreases apace. Hence, there is an optimum wall thickness for the fixed inner edge length and height of Au triangular nanoboxes.

Several observations can be made from the sensitivity parameters calculated in present study when compared with the previous studies on the diverse shape of Au nanoparticles and listed in Table 1 like: **i)** The refractive index sensitivity of nanoparticles enhanced significantly as we change the shape of the particle from spherical to sharp edge and can be further enhanced by making them hollow. **ii)** The figure of merit (FOM) of triangular nanoprisms is much higher than the nanospheres, nanocubes, and nanorods i.e., again the overall sensitivity response for the nanoparticles increases significantly by changing the shape towards the sharp edges. **iii)** FOM can be enhanced by changing the shape from solid to hollow i.e., from triangular nanoprisms to triangular nanoboxes. Au triangular nanoboxes should therefore be more advantageous than the other shapes when single nanoparticle is used for the refractive index change based plasmonic sensing.

**CONCLUSIONS**

We have simulated the extinction spectra of gold triangular nanoprisms and nanoboxes of various edge lengths, heights, and wall thicknesses using FDTD. From the spectra, we have shown that the refractive index sensitivity increases linearly and near-exponentially as the aspect ratio of nanoprisms increases and wall thickness of nanoboxes decreases, respectively. A figure of merit calculation shows that there is an optimum edge length in NIR region for triangular nanoprisms. The optimum wall thickness for various edge lengths and heights of the nanoboxes has turned out to be a promising candidate for their effective sensing applications. We also found that the higher refractive index sensitivity and figure of merit in triangular nanoboxes compared to the cubic nanoboxes and other solid structures is an outcome that depends on the nanoparticle shape.

**Figure Captions**

**Fig. 1:** FDTD simulated optical extinction spectra of Au nanoprisms of various edge lengths and heights. The surrounding refractive index is 1.33.

**Fig. 2: A)** Resonance wavelength (black squares) and refractive index sensitivity (red circles) plotted against aspect ratio of the nanoprisms. The straight lines represent the linear fit to the calculated data points (see text for linear fit Equation). **B)** Sensing figure of merit (FOM) plotted against the aspect ratio of triangular nanoprisms

**Fig. 3:** Refractive index sensitivity (black squares) and FOM (blue circles) plotted against the resonance wavelength. The straight line (in red) for the refractive index sensitivity shows the linear fit to the calculated data points

**Fig. 4:** FDTD calculated extinction spectra of **A)** triangular nanoboxes of inner edge length of 51.9 nm and inner height of 15 nm for various wall thicknesses ranging from 2.5 nm to 10 nm. The extinction spectrum for triangular prism of edge length of 51.9 nm and height of 15 nm is also shown (marked as **a**) and **B)** triangular nanoboxes of t = 5.0 nm for various inner edge lengths and heights. The surrounding refractive index is 1.33.

**Fig. 5:** The calculated **A)** refractive index sensitivity and **B)** FOM plotted against the wall thickness t.

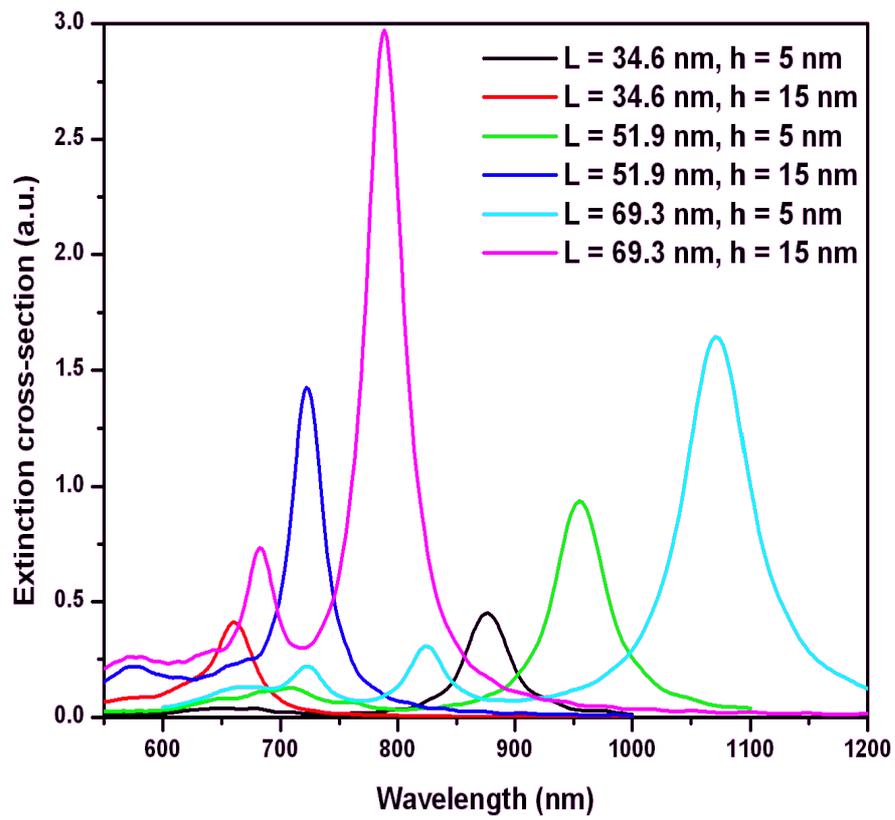

Figure 1

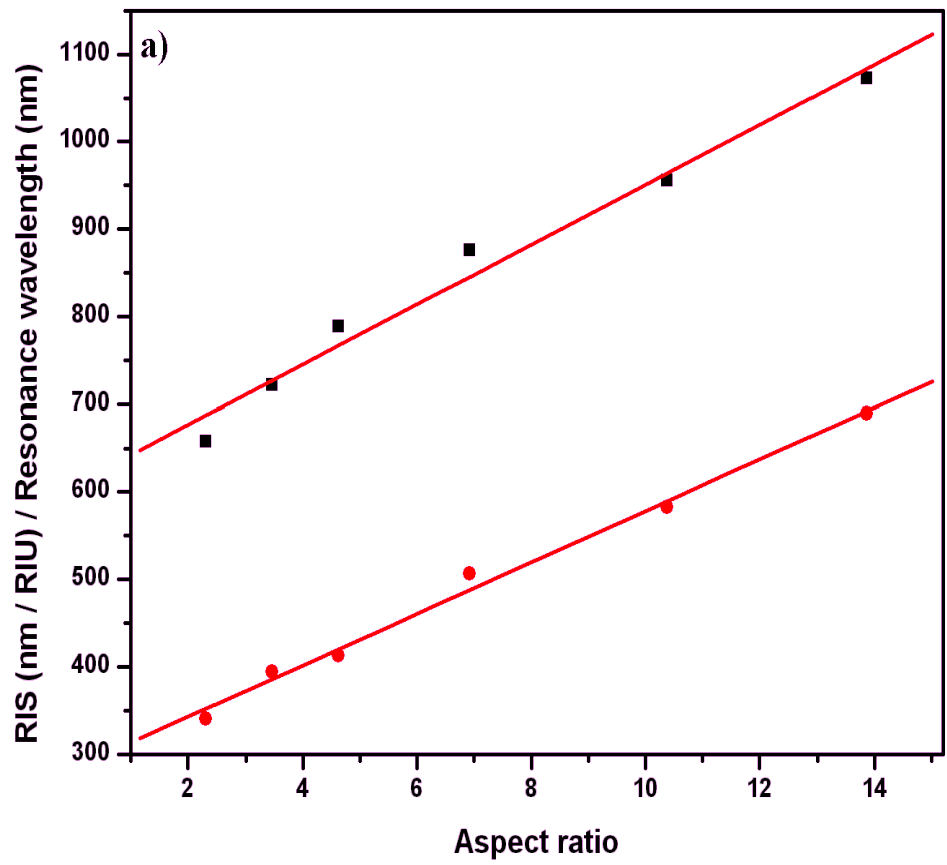

Figure 2a

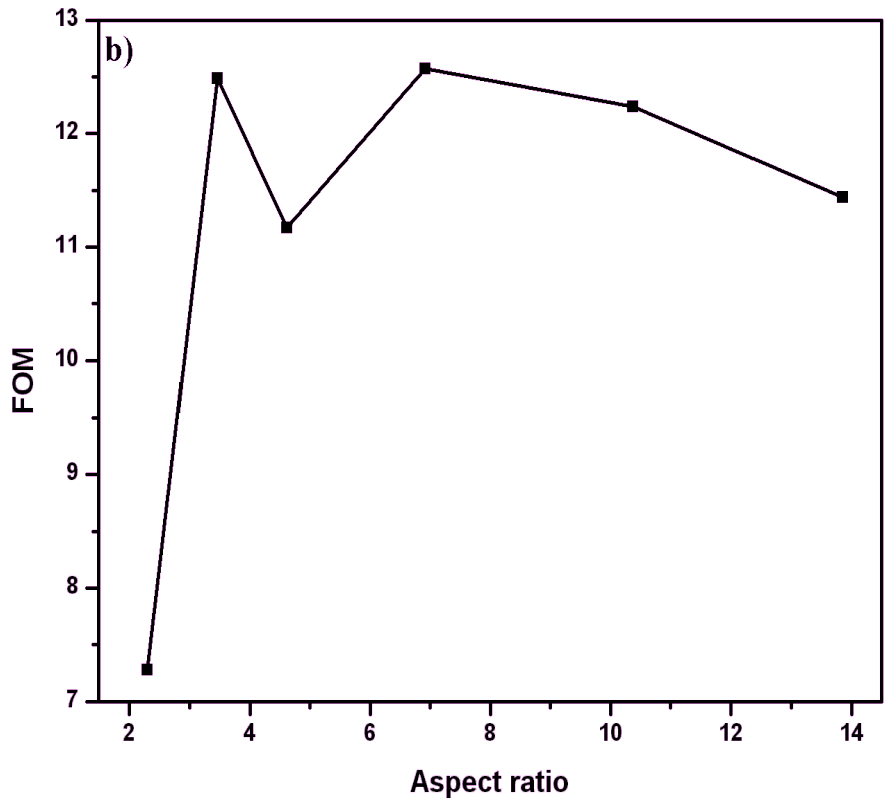

Figure 2b

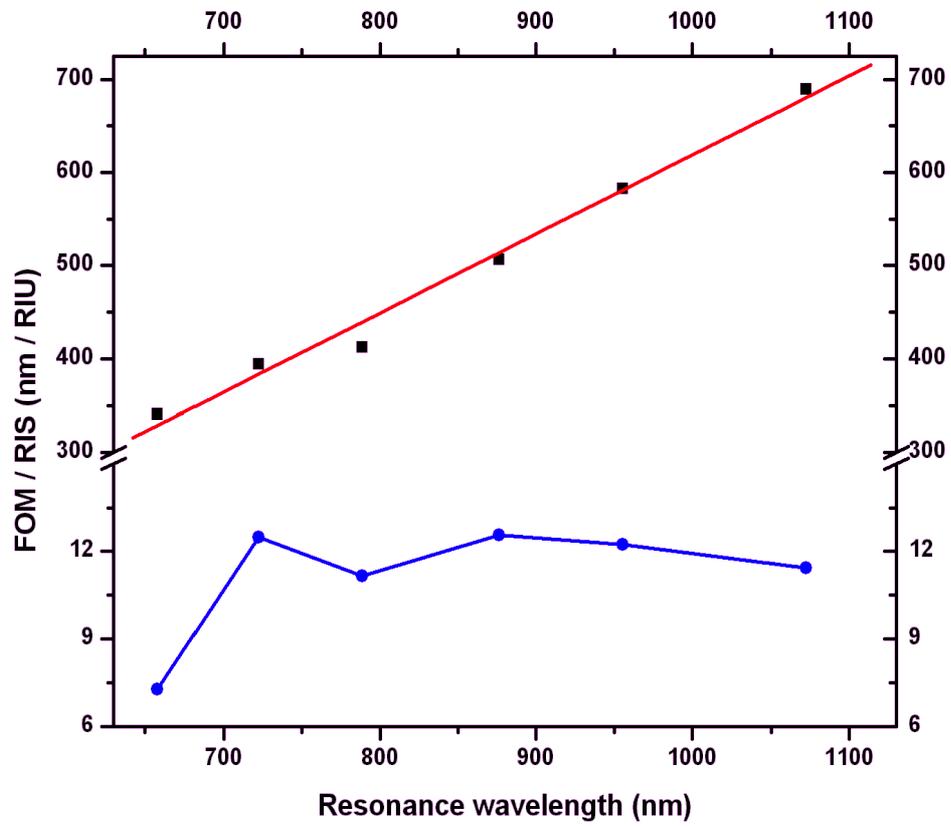

Figure 3

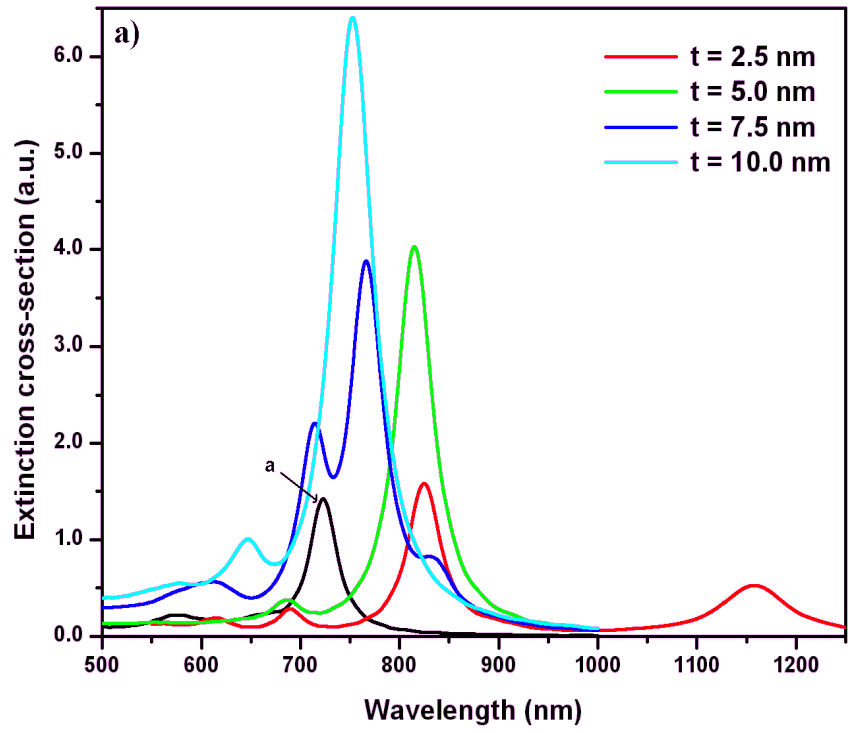

Figure 4a

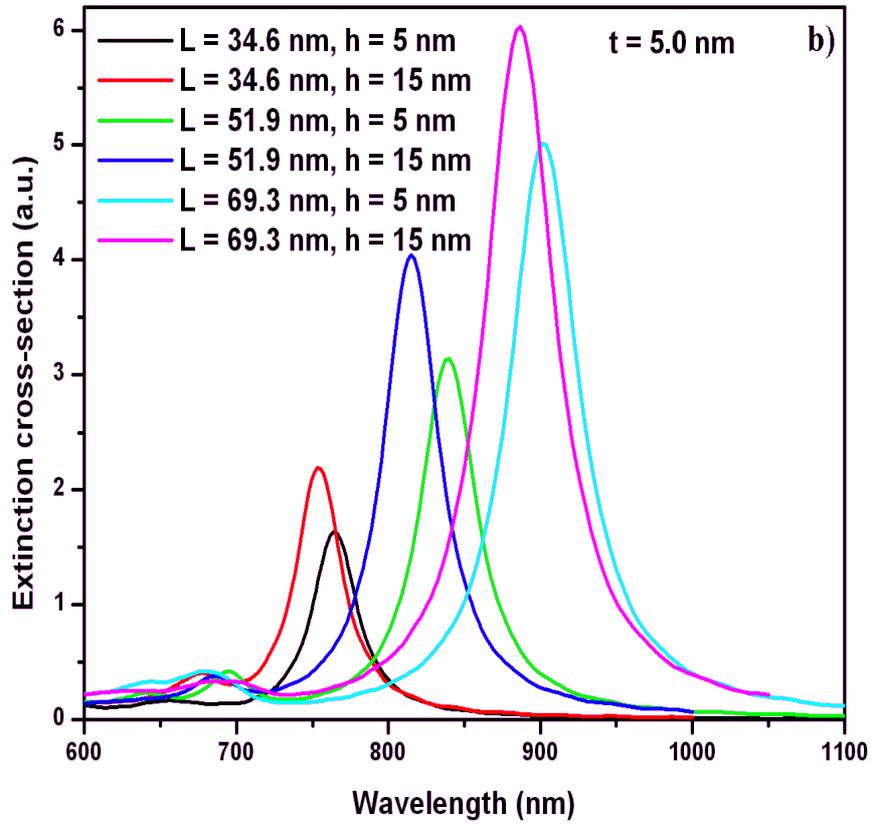

Figure 4b

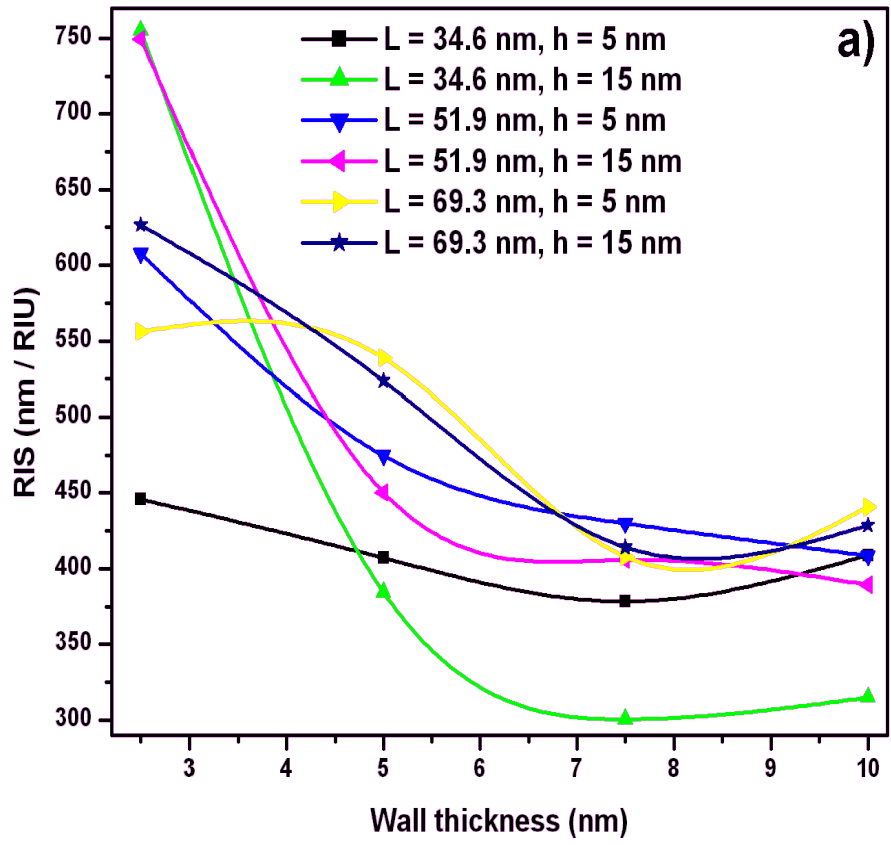

Figure 5a

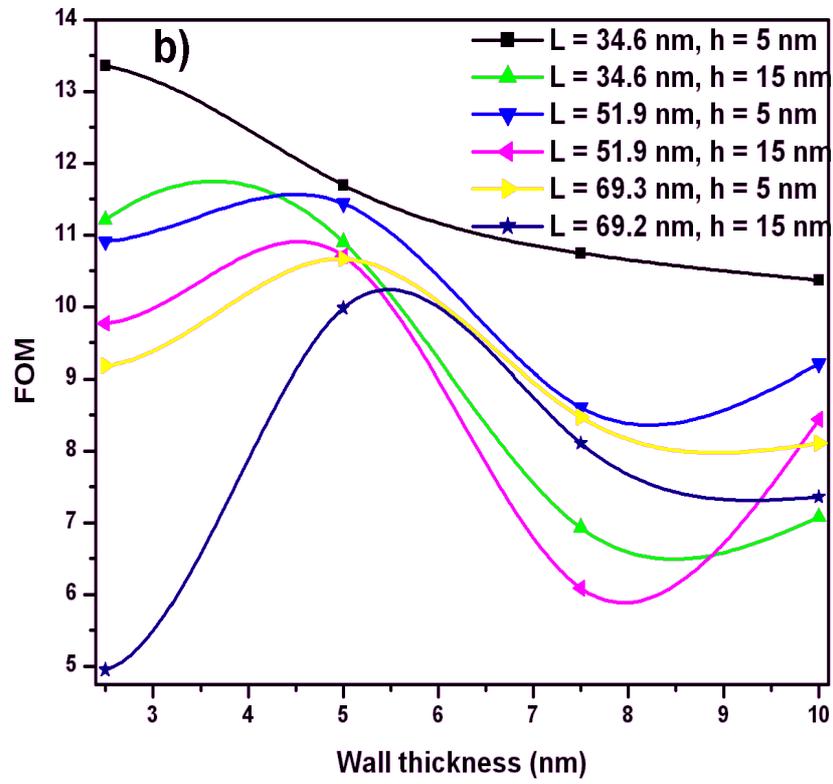

Figure 5b

**Table 1** Refractive index sensitivity and FOM of several single Au nanostructures. [a]for the triangular nanoboxes with inner edge length of 34.6 nm and height 5 nm. [b]for the cubic nanoboxes with an inner edge length of 30 nm and wall thickness of 5.7 nm. [c]for the nanorods width of 15 nm and aspect ratio of 2. [d]for the nanocubes of edge length 60 nm. [e]for the nanospheres of radius 40 nm.

| Au Nanoparticle structure | RIS (nm/RIU) | FOM | Ref. |
|---|---|---|---|
| Nanospheres[e] | 170 | 1.56 | 14 |
| Cubic nanobox[b] | 306 | 2.4 | 3 |
| Nanocubes[d] | 276 | 2.4 | 15 |
| Nanorods[c] | 510 | 3.8 | 17 |
| Nanoprisms | 394 | 12.6 | |
| Triangular nanoboxes[a] | 450 | 13.4 | |